\documentclass[journal]{IEEEtran}

\usepackage[colorinlistoftodos,prependcaption,textsize=tiny]{todonotes}
\usepackage[utf8]{inputenc}
\usepackage{epsfig}
\usepackage{graphicx}
\usepackage{amsmath}
\usepackage{amssymb}
\usepackage{color}
\usepackage{caption}
\usepackage{subcaption}
\usepackage{mathtools}
\usepackage{cite}

\usepackage{cite}

\ifCLASSINFOpdf
\else
\fi
\usepackage{url}

\begin{document}

\title{Optimal \emph{Dithering} Configuration Mitigating Rayleigh-Backscattering-Induced Distortion in Radioastronomic Optical Fiber Systems}

\author{Jacopo~Nanni, 
		Andrea~Giovannini, 
		Enrico~Lenzi, 
		Simone~Rusticelli,
		Randall Wayth  ˜\IEEEmembership{Member, ˜IEEE,}\\
		Federico~Perini, 
		Jader~Monari, 
		Giovanni~Tartarini  ˜\IEEEmembership{Member, ˜IEEE} 
\thanks{J.~Nanni, A.~Giovannini and G.~Tartarini are with the Dipartimento di Ingegneria dell'Energia Elettrica e dell'Informazione ``Guglielmo Marconi'', Universit\`a di Bologna, 40136 Bologna (BO), Italy (e-mail: jacopo.nanni3@unibo.it; andrea.giovannini14@unibo.it; giovanni.tartarini@unibo.it).}
\thanks{S.~Rusticelli is with 3PSystem s.r.l, via Riviera delBrenta 170, 30032 Fiesso d'Artico (VE), Italy (e-mail:s.rusticelli@3psystem.net).}
\thanks{F.~Perini, J.~Monari are with Institute of Radio Astronomy, National Institute for Astrophysics, Via Fiorentina 3513, 40059 Medicina (BO), Italy (e-mail: f.perini@ira.inaf.it;j.monari@ira.inaf.it;rusticel@ira.inaf.it).}
\thanks{E.~Lenzi is with RF Optics di Enrico Lenzi, Via Sopra Castello 21, 40061, Minerbio (BO), Italy (e-mail:enrico@enricolenzi.it).}
\thanks{R.~Wayth is with ICRAR/Curtin University, Bentley
Western Australia, 6102 (e-mail:R.Wayth@curtin.edu.au).}}


\maketitle

\begin{abstract}
In the context of Radioastronomic applications where the Analog Radio-over-Fiber technology is used for the antenna downlink, detrimental nonlinearity effects arise because of the interference between the forward signal generated by the laser and the Rayleigh backscattered one which is re-forwarded by the laser itself toward the photodetector.

The adoption of the so called \emph{dithering} technique, which involves the direct modulation of the laser with a sinusoidal tone and takes advantage of the laser chirping phenomenon, has been proved to reduce such Rayleigh Back Scattering - induced nonlinearities. The frequency and the amplitude of the \emph{dithering tone} should both be as low as possible, in order to avoid undesired collateral effects on the received spectrum as well as keep at low levels the global energy consumption. 

Through a comprehensive analysis of \emph{dithered} Radio over Fiber systems, it is demonstrated that a progressive reduction of the \emph{dithering tone} frequency affects in a peculiar fashion both the chirping characteristics of the field emitted by the laser and the spectrum pattern of the received signal at the fiber end.  

Accounting for the concurrent effects caused by such phenomena, optimal operating conditions are identified for the  implementation of the \emph{dithering tone}  technique in radioastronomic systems.
\end{abstract}


\section{Introduction}
\IEEEPARstart{S}{ystems} which adopt the Radio-over Fiber (RoF) technology in one of its most cost-efficient versions consist essentially in a radiofrequency (RF) signal which performs the direct intensity modulation (D-IM) of a laser source, propagates in a span of Standard Single Mode Fiber (SSMF) and is directly detected (DD) {at the receiver end} (D-IMDD RoF systems) \cite{Dat_Kanno,JQE19,MWP18}.

When the described scheme is utilized within Radioastronomic scenarios, peculiar features have to be considered, which distinguish these systems e.g. from  those which apply the D-IMDD RoF technology to the mobile network.

One of them is related to the frequencies of the transmitted RF signals, which in the case of the mobile network (4G, 5G signals and beyond) can reach several tens of GHz, and do not fall below around 700MHz\cite{MWP18,JQE19}, while within Radioastronomic application can also belong to intervals ranging from few tens  to few hundreds MHz \cite{SKA_LOW,AA2021}.

In addition to this, while RoF systems for the mobile network are typically designed assuming that at the optical transmitter's side the RF powers of the signals range roughly in an interval of $\pm$ few units of dBm, in the case of signals received by radioastronomic antennas these powers can exhibit values of just -60dBm  when they come from astronomical sources, while they arrive to a maximum of around -20dBm when they consist in undesired Radio Frequency Interference (RFI) signals coming from satellites, radio and/or television transmitters, etc. \cite{SPIE}.

As a consequence of the low power levels of the signals which directly modulate the laser source, in many RoF D-IMDD systems  utilized for radioastronomic applications it is relatively straightforward to reduce to negligible values the nonlinearities which typically bring detriment to RoF systems in the mobile networks. 
Indeed,  the spurious frequencies caused by the nonlinearity of the laser \emph{Power-Current} \cite{Petermann_book} curves typically in this case fall below the noise floor. Moreover, it is enough to operate in the vicinity of the second optical window (around the wavelength $\lambda = 1310nm$), to guarantee that, due to the low level of the RF power modulating the laser, the creation of nonlinearities generated by the combined effect of laser chirp and fiber chromatic dispersion is avoided \cite{Meslener}.

It has however been shown that in this kind of systems a primary cause of undesired nonlinear effects takes place when a generic RFI signal (with  frequency and angular frequency given respectively by $f_{RF}$ and $\omega_{RF}=2 \pi f_{RF}$) is transmitted in the RoF System together with the signals coming from sky sources, after that they have all been received by the Radioastronomic antenna \cite{ICTON_SKA}. Indeed, the portion of such RFI signal which is reflected by \emph{Rayleigh Backscattering} (RB) reaches the laser section, is partly reflected again and interacts {at the receiver end} with the transmitted RFI signal itself, generating nonlinear frequency terms (e.g. at frequencies $2f_{RF},\, 3f_{RF},...$.). Although the RFI disturbance at frequency $f_{RF}$ can be filtered out at the signal post-processing stage, the same cannot be done for all generated spurious frequency terms. 

{One countermeasure to the described problem could be represented by the introduction of a further optical
isolator at the fiber input section, which would be added to the one embedded in
the laser. However this solution would hardly meet the cost limitations (e.g. possibly 100{\it USD} or less for a single front-end receiver) imposed by the very high number of RoF downlinks which should be realized within  radioastronomic facilities operating at frequencies of tens/hundreds of MHz \cite{SKA_LOW, CHIME, GMRT, HERA, MWA, OVRO_LWA}.}

{In a previous work, an appropriate cost-effective solution has been demonstrated, consisting in the introduction of a \emph{dithering tone} as a further modulation current for the optical transmitter, which pushes the level of the nonlinear terms below the noise floor\cite{MWP19,JLT20}.}

{This solution had actually already been proposed in the past, within telecommunication scenarios, with the aim to reduce noise effects in optical fiber systems \cite{Lazaro}, \cite{Pepeljugoski}, and its application finalized to the reduction of the spurious terms constitutes an additional newly-evidenced advantage offered by this technique.}

In the application of the mentioned \emph{dithering tone} technique it would however be desirable that both the frequency $f_{d}$ (or angular frequency $\omega_{d}=2 \pi f_{d}$) and the amplitude $I_{d}$  of the current tone exhibit the lowest possible values.  The first requirement is related to the fact that the distance from $f_{RF}$ of the possibly generated spurious frequency terms, $f_{RF} \pm f_{d}, \, f_{RF} \pm 2f_{d},...$ should be lower than the resolution bandwidth of the reception filter, so that they can be removed from the received signal together with the RFI term at frequency $f_{RF}$. The second requirement comes instead from the necessity to keep at low levels the energy consumption of the tone generator.

To assist the designer in identifying the appropriate solution to the problem, a theoretical model of the laser behavior under current modulation in presence of optical feedback by RB is necessary. However, the models developed so far describing the effects produced by optical feedback under laser current modulation, are not referred to RB, since they consider only the reflection coming from a single external cavity \cite{Kobayashi,Helms_non_lin,Kikushima}, typically put on the back-facet of the laser. At the same time, the effects produced by RB feedback have been considered only with reference to the linewidth reduction \cite{Mark_LW} and optical frequency shift and hopping \cite{Mark,Chraplyvy}, in all cases for continuous-wave operation, without considering the presence of a modulating current.

In the present paper, a rigorous mathematical model based on the laser Rate Equations is developed to describe the effects of the feedback due to \emph{Rayleigh Backscattering} on an optical field modulated by an RF signal. The model, which will be shown to be in agreement with experimental measurements, is utilized to describe the effects of the introduction of the  \emph{dithering tone} and to identify the optimal working point which allows keep both $I_{d}$  and $\omega_{d}$ as low as possible.

%

The paper is organized as follows: Section II provides a description of the model developed to study the impact of RB on the characteristics of the field emitted by a directly modulated laser. Section III describes the experimental results, which confirm the correctness of the theoretical model adopted and allow to identify the optimal operating conditions for the  implementation of the \emph{dithering tone}  technique. Finally, conclusions will be drawn.

\section{Theoretical Model}

\begin{figure}[t]
\centering
\includegraphics[scale=0.26]{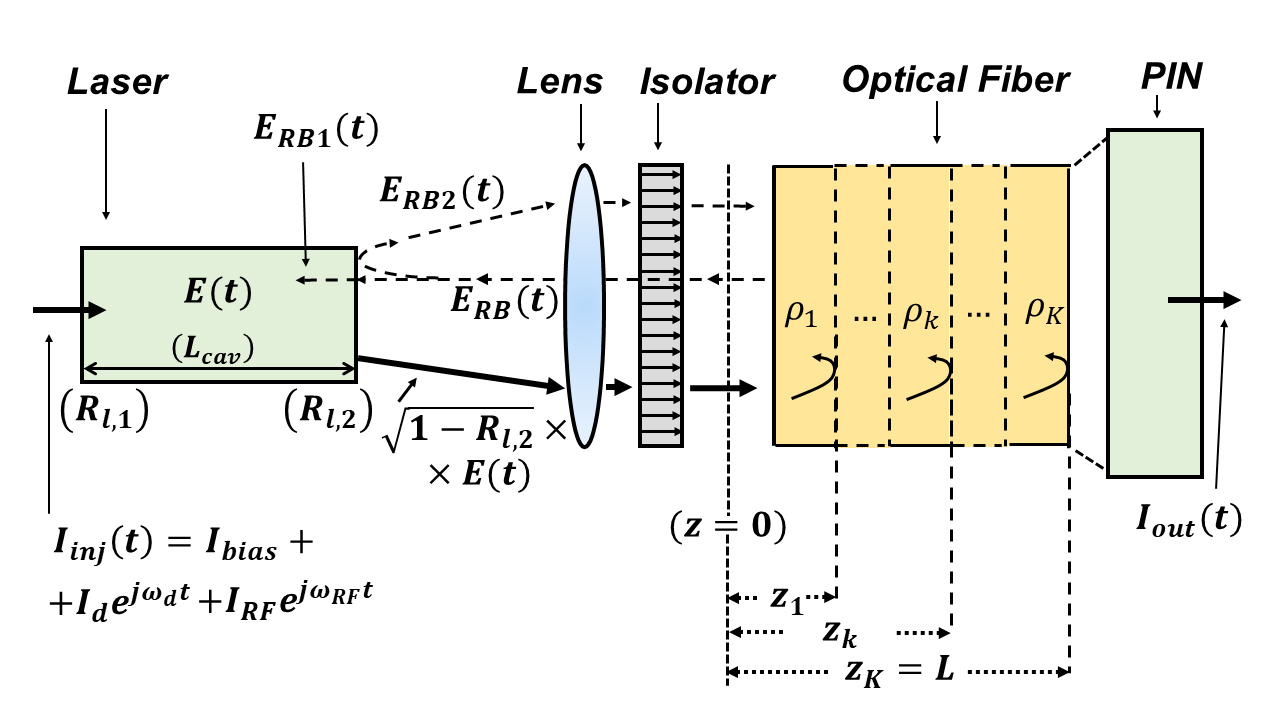}
\caption{Scheme of the structure considered.}
\label{fig:initial_scheme}
\end{figure}

The theoretical model presented in this work is based on the physical structure shown in Figure \ref{fig:initial_scheme}, composed by a laser-based optical transmitter whose emitted field is coupled  through a focusing lens and an isolator to an optical fiber and reaches finally an optical receiver based on a PIN photodetector. 
The laser cavity, exhibits a length $L_{cav}$ and power reflectivities at the facets $R_{l,1}$ and $R_{l,2}$, respectively. {The injection current $I_{inj}(t)$ can be written in complex notation as $I_{inj}(t)= I_{bias}+ \sum_{i} I_{i}e^{j\omega_{i}t}$, where $I_{bias}$ represents the bias current, and $I_{i}$ represents the amplitude of the generic $i-th$ sinusoidal RF tone which directly modulates the laser, with angular frequency $\omega_{i}$ and frequency $f_i=\omega_i/(2 \pi)$. For formal simplicity and without loss of generality, in the following only two modulating tones will be considered, namely a term with amplitude $I_{RF}$ and angular frequency $\omega_{RF}=2 \pi f_{RF}$, and the so called  {\it dithering tone} with amplitude  $I_{d}$ and angular frequency $\omega_{d}=2 \pi f_{d}$, resulting in:
\begin{equation}
I_{inj}(t)=I_{bias}+I_{d}e^{j\omega_{d}t}+  I_{RF}e^{j\omega_{RF}t}
\label{eq_I_inj}
\end{equation} }
For this reason, instead of $\sum_{i= d, RF} \dots$, the simplified notation $\sum_i \dots$ will be adopted throughout the paper.

{In principle, the electrical field emitted by the laser source and propagating into the fiber should be represented in vector form, to account for the evolution of its state of polarization (SOP). However, the phenomenon of Rayleigh backscattering, which takes place within the fiber in correspondence to the reflection sections $z_k$, $k=1,\dots, K$ (see again Figure \ref{fig:initial_scheme}) can be regarded as substantially isotropic, while the change of the SOP of the field between $z_k$ and an adjacent section can be described by an appropriate Jones matrix, which is unitary.
Consequently, it can be shown that for a given initial SOP and in absence of Polarization Dependent Losses (PDL), the electrical field exhibits at the fiber end section a corresponding final SOP on which the quantities which are of interest in the present work (e.g. powers received at different frequencies) do not depend \cite{Gysel_JLT,Kapron}.}
 
{In presence of Polarization Dependent Losses (PDL), the quantities which are of interest in the present work can instead exhibit variations depending on the final SOP of the electric field.
This could be appreciated for example in presence of changes of environmental quantities (e.g. temperature). Indeed, while it can be reasonably assumed that the devices located within the transmitter (single-mode laser, lens, isolator) still determine a stable SOP for the emitted field, the optical fiber can undergo local birefringence perturbations which can change in real time the SOP of the field impinging the detecting area of the PIN. In presence of PDL of this last device, the received powers would then exhibit a polarization-dependent fluctuating behavior. However the relative variations of the measured quantities can be typically considered as low (e.g.few tenths of $dB$ or less), and, at the same time, on fiber connections of the order of few km length, like the ones considered in this work, they can be appreciated on time scales (e.g. tens of seconds or more) which are much longer than the ones of the phenomena investigated\cite{SPIE}.}

{For all these reasons, in the rest of the paper the electrical field will be described as a scalar quantity without losing the generality of the results that will be obtained.}

\subsection{Re-transmitted and re-injected portions of the Rayleigh Backscattered field}
\label{sub:RB1_RB2}
{By virtue of the considerations developed above, t}he electrical field  generated inside the laser can be expressed as 

\begin{equation} 
E(t)=|{E}(t)|e^{j\left[ \omega_{0}t+\theta(t)\right]} e^{j\phi(t)}
\label{eq:laser_field}
\end{equation}

where  $|{E}(t)|$ is its module, $\omega_{0}$ the optical angular frequency that the field would exhibit in absence of modulating tones and in absence of RB as well, $\theta(t)$ is the time-varying phase due to their presence, while $\phi(t)$ represents the laser phase noise contribution. The expressions of $|E(t)|$ and $\theta(t)$  are given by:

\begin{eqnarray}
&&|E(t)|= E_0\sqrt{1+ \zeta_{RB}+\sum_i m_icos(\omega_it)}
\label{eq:E}\\
&&\theta(t)=\Delta \omega t\,\, +\sum_{i} |M_i|cos(\omega_{i}t+ \angle M_i)
\label{eq:theta}
\end{eqnarray} 

In \eqref{eq:E} $E_0 =\sqrt{\eta (I_{bias}-I_{th})}$ is the electrical field amplitude, with $I_{th}$ threshold current and $\eta$ power-current slope efficiency of the laser, $\zeta_{RB}$ represents a small increment to the unit number under square root, which derives from the presence of RB and will be further described in the next subsection, while $m_i=I_i/(I_{bias}-I_{th})$ represents the optical modulation index at the considered frequency.

In \eqref{eq:theta} $\Delta \omega$ represents a variation of the angular frequency due to RB and to direct modulation, whose presence determines the total optical angular frequency of the emitted field to be $\omega_{opt}=\omega_{0}+ \Delta \omega$. Moreover, the quantities $|M_i|$ and $\angle M_i$ are respectively module and phase of the phase modulation index due to the chirp effect at  $\omega_i$ \cite{Nanni_AO}. 

In \eqref{eq:laser_field} the quantity $e^{j\phi(t)}$ is a random function whose Power Spectral Density, determined as Fourier Transform of its autocorrelation function, is given by: 
\begin{equation}
\mathcal{F_{\xi}} \left\{ \left\langle  e^{j\phi(t)-\phi(t-\xi)}
            \right \rangle \right\}=           
\mathcal{F_{\xi}} \left\{ e^{-|\xi|\Delta_{\Omega}} \right\}
=\frac{2\Delta_{\Omega}}{\Delta_{\Omega}^2+\omega^2}
\label{eq:phase_noise}
\end{equation}
where $\mathcal{F_{\xi}}\{\cdots\}$ , $\left \langle \cdots \right \rangle$ indicate respectively the operations of Fourier Transformation with respect to the variable $\xi$, and time averaging. The quantity $\Delta_{\Omega}=1/\tau_{coh}$ is the coherence angular frequency band of the laser source, reciprocal of its coherence time $\tau_{coh}$ \cite{Gallion}.

Still referring to Fig. \ref{fig:initial_scheme}, the field given by \eqref{eq:laser_field} exits the laser right-hand-side facet, is focused by a coupling lens into an isolator, and is subsequently injected at the input section ($z=0$) of the optical fiber of length $L$, propagating  with a group velocity $v_g$. 
Putting $\tau=z/v_g$, at the generic section $z \in[0,L]$ of the fiber the resulting electric field can then be written as:

\begin{eqnarray}
&&E_{TX}(t,z) =\sqrt{1-R_{l,2}}\cdot \sqrt{\eta_c} \cdot {{E}}(t-\tau) e^{-\alpha v_g \tau}
\label{eq:fiber_field}
\end{eqnarray}

{where  $\eta_c$ is the power coupling coefficient between the laser and the optical fiber given by the coupling lens, while $\alpha$ is the attenuation constant of the fundamental mode of the optical fiber.}

As {above} mentioned, as $E_{TX}$ propagates along the fiber, the phenomenon of Rayleigh backscattering takes place. The inhomogeneity of the longitudinal profile of the refractive index of the optical fiber determines indeed the presence of many reflection points at the coordinates $z_k, \,\,k=1,\dots, K$, each one of which generates a back reflected electrical field. At the output facet  of the laser cavity a Rayleigh Backscattered field is then present, given by:

\begin{equation}\label{eq:RB field of}
E_{RB}(t) =\frac{\sqrt{1-R_{l,2}}\cdot \eta_c}{\sqrt{\alpha_{iso}}}\sum_{k=1}^K \rho _k {{E}}(t-2 \tau _k)e^{-\alpha v_g 2 \tau _k}
\end{equation}
where $\alpha_{iso}$ is the isolator power attenuation in the backward direction, while $\tau_k = z_k/v_g$. Then,  {$\rho_k=|\rho_k|e^{j \angle \rho_k}$} is the complex reflection coefficient at the $k-th$ section due to RB. Note that the \textit{weak-feedback} approximation, i.e. $|\rho_k|<<1$, is taken, meaning that possible multiple reflections within a single fiber section are not considered.

The different $\rho_k$'s are assumed to be complex random variables, with zero mean value and Gaussian distribution. The variance of both their real and imaginary parts is $\sigma^2_{\rho}=\frac{1}{2} \alpha_s \mathcal{S} d z_k$, where  $\mathcal{S}$ is the so-called \emph{backscattering factor} or \emph{recapture factor}, which depends on the characteristics of the fiber considered and exhibits typical values of the order of $10^{-3}$ in case of the standard G.652 fiber, $\alpha_s$ is the Rayleigh attenuation coefficient, which for the considered wavelengths can be assumed to coincide with $\alpha$, and $dz_k=z_k - z_{k-1}$ is the length of the interval considered  \cite{Brinkmeyer}. The $\rho_k$'s are also assumed to be {\it delta-correlated}, namely:

\begin{equation}
{\mathbb{E}[ \rho_k \rho^{\ast}_h] }= \alpha_s\mathcal{S}\,dz_k \delta_{kh}
\label{eq:rho_k}  
\end{equation}

\noindent where $\mathbb{E}[\cdot]$ represents the expected value operator, while $\delta_{kh}$ is the {\it delta Kronecker function} \cite{Gysel_JLT}.

Still referring to Fig. \ref{fig:initial_scheme}, a portion of the field reported in \eqref{eq:RB field of}, given by $E_{RB1}(t)=\sqrt{1-R_{l,2}}E_{RB}(t)$, is re-injected into the laser cavity. {As will be shown in Subsection \ref{sub:rate_eq}, this fact has significant consequences on the phase modulation index due to the chirp effect $M_d$ at $\omega_d$, which also influences the identification of the optimal frequency and amplitude characteristics of the {\it dithering tone} itself}. 

Another portion of this field, given by $E_{RB2}(t,z)=\sqrt{R_{l,2}}E_{RB}(t,z)$ is instead reflected by the laser facet. As will be shown in  Subsection \ref{sub:final_spectrum}, this reflected field adds to the one given by \eqref{eq:fiber_field} in input to the optical fiber, contributing to the final form of the power spectral density of the photocurrent detected at the receiving end of the system. 
 The knowledge of the behavior of this quantity allows to put into evidence particular physical aspects which also contribute to determine the optimal frequency and amplitude characteristics of the {\it dithering tone} to be utilized.


\subsection{Re-injected portion of the RB field and effect on the Chirp term}
\label{sub:rate_eq}
{The Rate Equations, expressed below in the so called Semi Classical form \cite{Petermann_book}  \cite{Coldren} \cite{Agrawal_Dutta}, describe the behaviors of $S(t)=|E(t)|^2$, $\theta(t)$ and $N(t)$, which represent respectively the number of photons in the laser cavity, the previously introduced time varying phase $\theta(t)$ and the carrier density $N(t)$, expressed in  $\left[\frac{1}{m^3}\right]$:}

\begin{align}
\frac{d S(t)}{dt}&=\left(G(N,S)-\frac{1}{\tau_p}\right)S(t)+ \Gamma_S(t)\label{eq:S}\\
\nonumber\\
\frac{d\theta(t)}{dt}&=\frac{\alpha_H}{2}G_N(N(t)-N_{th})-\Gamma_{\theta}(t)\label{eq:phi}\\
\nonumber\\
\frac{dN(t)}{dt} &= \frac{I_{inj}(t)}{eV}-\frac{N(t)}{\tau_s}-\frac{G(N,S)}{V}S(t)\label{eq:N}
\end{align} 

{At the second side of (\eqref{eq:S}) the quantity $G(N,S)$ is the laser cavity gain which depends both from $N$, and  $S$ through the relation:
\begin{equation}\label{eq:gain}
G(N,S) = \frac{g_0}{1+\epsilon S}(N-N_{tr})\simeq g_0\cdot(1-\epsilon S)(N-N_{tr})
\end{equation}}

{where the \emph{transparency level} $N_{tr}$ is the density of carriers in the cavity for which $G(N,S)=0$, $\epsilon<<1/S$ is the gain compression factor and $g_0$ represents the nominal gain-slope in absence of compression, multiplied by the volume of the laser cavity \cite{Coldren}. The quantity $\tau_p$ represents the average photon's lifetime, or inverse of the cavity losses.}

{The term $\Gamma_S(t)$ at the right-hand-side of \eqref{eq:S}, which  represents the RB contribution to $dS(t)/dt$, is given by:
\begin{eqnarray}
&&\Gamma_S(t)=2\sum_{k=1}^K|C_k| \sqrt{S(t)S(t-2\tau_k)}\cdot\nonumber\\
&&\cdot\cos[\omega_{0}\cdot 2\tau_k +\theta(t)-\theta(t-2\tau_k)-\angle \rho_k]
\label{eq:Gamma_S}
\end{eqnarray}
and will be derived in Appendix \ref{app:1}. Within \eqref{eq:Gamma_S}, $C_k$ is the $k$-th compound-cavity coefficient \cite{Kobayashi} defined as follows:
\begin{equation}
C_k=\frac{(1-R_{l,2})\eta_c}{\sqrt{\alpha_{iso}}}\rho_ke^{-\alpha v_g 2 \tau_k}\frac{1}{{2}\tau_{cav}}\frac{1}{\sqrt{R_{l,2}}}
\end{equation}
where $\tau_{cav}=2\cdot n_{cav}L_{cav}/c$ is the round-trip time of the laser cavity with refractive index $n_{cav}$.}

{By solving the laser rate equations in the reference steady state case where both external RF modulation and RB feedback are absent, it results that the carrier density is equal to the amount $N_{th}$ which balances the cavity losses, and it results also that number of photons is $S_0= \frac{\tau_p}{e}\left(I_{bias}-I_{th}\right)$ with $I_{th}=N_{th}\frac{eV}{\tau_s}$, where the relationship ${\tau_p} = 1/G(N_{th},{S}_{0})$ has been exploited.}

{At the second side of (\ref{eq:phi}) $\alpha_H$ is the linewidth enhancement factor of the laser source \cite{Coldren} while it is $G_N = \left. \frac{\partial G}{\partial N}\right|_{S_0}=g_0(1-\epsilon S_0)$. 

Similarly to (\ref{eq:S}), the last term at the right-hand-side of (\ref{eq:phi})  represents the RB contribution to $d \theta(t)/dt$:
\begin{eqnarray}
&&\Gamma_{\theta}=\sum_{k=1}^K|C_k|\sqrt{\frac{S(t-2\tau_k)}{S(t)}}\cdot \nonumber \\
&&\sin[\omega_{0}\cdot 2\tau_k +\theta(t)-\theta(t-2\tau_k)-\angle \rho_k]
\label{eq:Gamma_PHI}
\end{eqnarray} 
and will also be derived in Appendix \ref{app:1}.}

{Finally, at the second side of (\ref{eq:N}) the quantities  $e$,  $V$, $\tau_s$ represent respectively the electron charge, the volume of the active region and  the carrier-lifetime}.

{Assuming that partial RB field re-injection and RF modulation of the injection current determine only a perturbative effect on the laser emission characteristics, the behavior of $G(N,S)$ can be represented by its expansion to the first order: 
\begin{equation}\label{eq:gain_approx}
G(N,S) \simeq  G(N_{th},{S}_{0})+G_N(N-N_{th})+G_S(S-S_{0})
\end{equation}
where   $G_S = \left. \frac{\partial G}{\partial S}\right|_{N_{th}}=-g_0 \epsilon (N_{th}-N_{tr})$.} 


%
%

{Within the hypothesis taken, in presence of an injection current $I_{inj}(t)$ given by the second side of (\ref{eq_I_inj})  $S(t)$, $\theta(t)$ and $N(t)$ exhibit expressions given by: 
\begin{align}
S(t) &= S_0 + \Delta S +\sum_{i} S_{i}e^{j\omega_{i}t}\label{eq:delta S}\\
\theta(t) &= \Delta \omega t + \sum_{i} M_{i}e^{j\omega_{i}t}\label{eq:delta theta}\\
N(t) &= N_{th} + \Delta N +\sum_{i} N_{i}e^{j\omega_{i}t}\label{eq:delta N}
\end{align}}
{\noindent where \eqref{eq:delta theta} is \eqref{eq:theta} re-written in complex notation and where 
$(\Delta S, 
\Delta \omega, \Delta N)$, 
and $(S_{i}, M_{i}, N_{i}) \, {\forall i}$, are the unknowns which, within the perturbative regime taken, respect the conditions:
\begin{eqnarray}
&&\Delta S, \, |S_{i}|_{\forall i} << S_0
\label{eq:perturb_1} \\
&&  \Delta \omega, \,|\omega_{i}M_{i}|_{\forall i}  << \omega_{0} 
\label{eq:perturb_2} \\
&&\Delta N, \,|N_{i}|_{\forall i}  << N_{th}
\label{eq:perturb_3}
\end{eqnarray}}
{In line with \eqref{eq:perturb_1}, \eqref{eq:perturb_2}, \eqref{eq:perturb_3}, it will be assumed that  the perturbation caused the generic $i-th$  modulating tone can be determined separately from the one due to the other, namely that the expression of the triplet ($S_{{i}}$, $M_{{i}}$, $N_{{i}}$) can be determined just considering $I_{inj}(t)=I_{bias}+ I_{{i}}e^{j\omega_{{i}}t}$ ($i=d$ or $i=RF$).}

{Inserting the relations \eqref{eq:delta S}, \eqref{eq:delta theta}, \eqref{eq:delta N} with the summations composed by one element  into \eqref{eq:S}, \eqref{eq:phi}, \eqref{eq:N}, and considering the terms at $DC$ and at $\omega_i$ of the Jacobi-Anger expansions of \eqref{eq:Gamma_S} and \eqref{eq:Gamma_PHI} (detailed in Appendix \ref{app:2}), it is possible to obtain two systems of three equations. The first one, which is omitted for brevity, represents the steady-state terms and allows to determine $\Delta S, \Delta \omega, \Delta N$, while the second one represents the modulating terms with angular frequency $\omega_i$, which allow to determine $S_{i}, M_{i}, N_{i}$.}

{The latter of the two systems assumes the following form:}
{
\begin{equation}
\overline{\overline{A}}\cdot\left(\begin{array}{c}
S_i\\
M_i\\
N_i
\end{array}\right)=\left(\begin{array}{c}
0\\
0\\
\frac{I_i}{eV}
\end{array}\right)
\end{equation}
\noindent where the matrix $\overline{\overline{A}}$ is given by:
\begin{align} 
&\overline{\overline{A}} =\label{eq:system}\\
&\left(\begin{array}{ccc}
j\omega_i-G_SS_0&+2S_0\gamma_{S,i}&-G_NS_0\\
\\
0&j\omega_i+\gamma_{\theta,i}&-\frac{\alpha_H}{2}G_N\\
\\
\frac{1}{\tau_pV}+\frac{G_S S_0}{V}&0&j\omega_i+\frac{1}{\tau_s}+\frac{G_N S_0}{V}
\nonumber
\end{array}\right)
\end{align}}

{and where the elements $\gamma_{S,i}$ and $\gamma_{\theta,i}$, located in positions $(1,2)$ and $(2,2)$, are derived from $\Gamma_S$ and $\Gamma_{\theta}$, as shown in Appendix \ref{app:2}.} 

As will be evidenced in the next Subsection, the higher values exhibited by the module $|M_d|$ of the phase modulation index due to the chirp effect at $\omega_d$,  the more effective the \emph{ dithering technique} results in reducing to negligible levels the undesired RB-related nonlinearities. For this reason, the behavior of $M_i$, which is obtained by solving \eqref{eq:system}, will be analyzed in the following.

\begin{table}[t]
\centering
\caption{{\it Typical Orders of Magnitude of Rate Equations Parameters.}}
\renewcommand\arraystretch{1.0}{
\begin{tabular}{c|c|c}
\textbf{Symbol}& \textbf{Physical meaning}& \textbf{Order of Magnitude}\\
\hline
$g_{0}$  &Gain slope & $\sim 10^{-12}$ $m^3/sec$\\
$N_{tr}$& Carriers Transparency Level& $\sim 1.\cdot 10^{24}$ $m^{-3}$\\
$N_{th}$& Carriers Threshold Level& $\sim 2.\cdot 10^{24}$ $m^{-3}$\\
$\epsilon$       & Gain compression factor        & $\sim 10^{-8}$\\
$\tau_p$  & Photon Lifetime                & $\sim 10^{-12} sec$\\
$\tau_s$		& Carrier Lifetime       &$\sim 10^{-9} sec$\\
$R_{l,1}, R_{l,2}$& Laser mirrors reflectivities &$\sim 0.3,\dots,0.5$\\
$L_{cav}$		& Laser cavity length				&$\sim few \, 100 \mu m$\\
$V$		& Volume of Laser active region			&$\sim 10^{-16} m^3$\\
$n_{cav}$		&Laser cavity refractive index       & $\sim 3.5,\dots,5$\\
$\alpha_H$		& Linewidth enhancement factor       & $\sim 3,\dots,6$\\
  &Number of photons in steady state & \\
$S_{0}$  &(assuming for $I_{bias}-I_{th}$ & $\sim 10^{5}$ \\
& a value of few tens  $mA$) & \\
$G_N$&$\partial G / \partial N$ in steady state& $\sim 10^{-12}$ $m^{3}$\\
$G_S$&$\partial G / \partial S$ in steady state& $\sim 10^{4}$ \\
\hline
\end{tabular}}\label{laser:parameters}
\end{table}


In doing so, some simplifications can be taken, which allow to put this quantity in a more readable form. 
Indeed, considering modulating frequencies $f_i$  which can range from $KHz$ to hundreds MHz, and assuming for $I_{bias}$ a value such that $I_{bias}-I_{th} \sim$ few  tens of $mA$, taking into account the typical orders of magnitude of Rate Equations parameters reported in Table \ref{laser:parameters}, the determinant of $\overline{\overline{A}}$ can be reduced to:

\begin{equation}
det \left[ \overline{\overline{A}} \right] 
\simeq \frac{G_N S_0}{\tau_p V} \left[j \omega_i+\gamma_{RB,i} \right]
\label{eq:det_M}
\end{equation}

where it has been put:

\begin{eqnarray}
&&\gamma_{RB,i}=\gamma_{S,i}-\alpha_H \gamma_{\theta,i}\simeq \sqrt{1+\alpha_H^2}\cdot \sum_{k=1}^{K} |C_k| \cdot \label{eq:gamma_RB}\\
&&\cdot \frac{2J_1(X_{i,k})}{X_{i,k}} \left(1-e^{-j\omega_i2\tau_k}\right)\cos\left(\psi_k-\tan^{-1}\alpha_H\right)
\nonumber
\end{eqnarray}

with:

\begin{eqnarray}
&&X_{i,k}=  2 |M_i|sin(\omega_i \tau_k +\angle M_i)
\label{eq:X_i_k}\\
&& \psi_k =\omega_{opt} 2\tau_k - \angle \rho_k
\label{eq:psi_k}
\end{eqnarray}

The resulting expected value of $|M_i|$ is then given by the expression:
\begin{equation}
|M_{i}|
= \mathbb{E}\left[
\frac{\frac{\tau_p}{e}\frac{\alpha_H}{2}G_S\,I_i}
{\sqrt{\omega_i^2 + |\gamma_{RB,i}|^2 }} \right]
\label{eq:theta_i}
\end{equation}
in which the statistical expectation can be straightforwardly applied through \eqref{eq:rho_k} and utilizing the properties of the $\rho_k$'s  specified in Section \ref{sub:RB1_RB2}. 
It can be at first noted that the value of $|M_i|$ are proportional to the amplitude of the modulating current $I_i$. Secondly, regarding the dependence of the frequency $f_i$ (or angular frequency $\omega_i$) of the modulating current, it can be first observed 
that in case the RB effects were absent, the value of $|M_i|$ would result inversely proportional to $\omega_i$. 

\begin{figure}[t]
\centering
\includegraphics[scale=0.33]{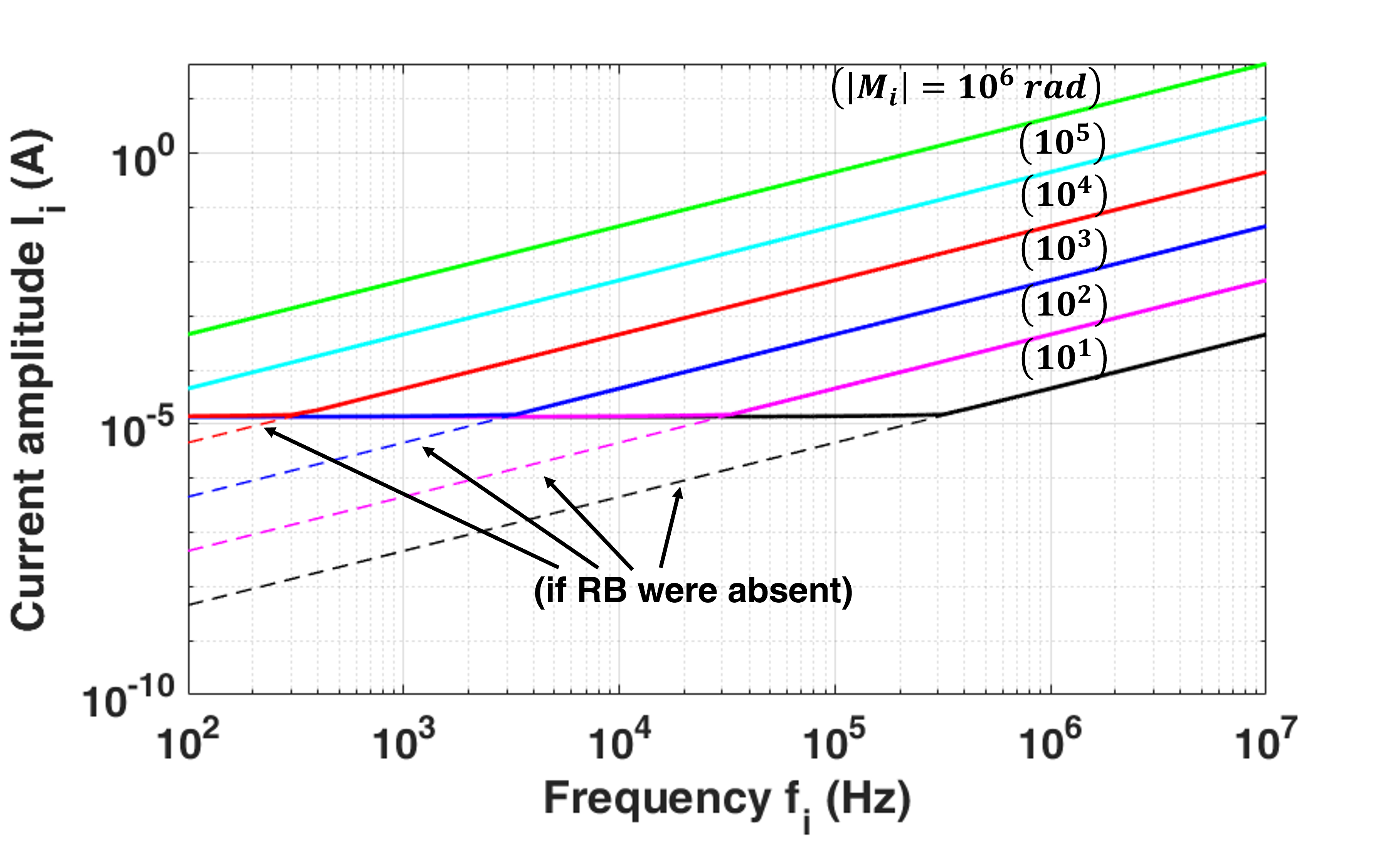}
\caption{Values of the modulating current amplitude $I_i$ which, for a given modulating frequency $f_i$ determine assigned values of the module of the phase modulation index due to the adiabatic chirp $|M_i|$. Continuous lines: realistic cases where the effect of RB is taken into account.  Dashed lines: theoretical values of $I_i$ in case no RB field re-injection were present. Note that the subscripts can be assumed as $i=RF$ or $i=d$.}
\label{fig:theta_i_module}
\end{figure}

In the considered case, the dependence of $|M_i|$ on $\omega_i$ does not instead follow such simple pattern. Solving \eqref{eq:theta_i} iteratively (since $|M_i|$ is contained in $\gamma_{RB,i}$) it is indeed possible to obtain the curves reported in Fig.\ref{fig:theta_i_module}, which have been modelled in accordance with the orders of magnitude of the parameters reported in Table \ref{laser:parameters} and Table \ref{tab:parameters}. 

At the right hand side of the figure it can be noted that for frequencies greater than few hundred MHz it can effectively be assumed for all the reported curves  $|M_i|\simeq \frac{K_{f,i}I_i}{f_i}$ where $K_{f,i}=\frac{1}{2 \pi}\frac{\tau_p}{e}\frac{\alpha_H}{2}G_S$, is the so called {\it adiabatic chirp factor}, expressed in $[Hz/A]$. Indeed, a constant value $|M_i|=10^n \, rad$ ($n=1, \dots, 6$ in the figure) corresponds to the linear relationship $I_i=\frac{10^n \, rad}{K_{f,i}}f_i$. This is the behavior of $|M_i|$ when $i=RF$ is considered, since the undesired RFI signals which affect radioastronomic plants operating in the lower bandwidth of the radioastronomic spectrum exhibit frequencies of tens to hundreds of MHz \cite{JLT20}. 

When instead the {\it dithering tone} is considered ($i=d$), it is however convenient to give lower values to $f_d$. For example, a given constant value $|M_d|=10^3\, rad$ would be obtained with $I_d\simeq 10$ mA if $f_d\simeq 2$ MHz were chosen, but can be obtained with $I_d\simeq 0.1$ mA if $f_d\simeq 20$ kHz. Note that giving low values to $f_d$ fulfills also the requirement to keep the terms $f_{RF} \pm f_{d}, \, f_{RF} \pm 2f_{d},...$ within the resolution bandwidth of the reception filter, as mentioned in the Introduction.

Still considering as example the curve referred to the value $|M_d|=10^3 \, rad$, it can however also be noted that for frequencies lower than few kHz the value of $I_d$ to be utilized becomes practically independent from $f_d$. This happens because, as can be shown with a detailed analysis of its composing terms, $|\gamma_{RB,d}|$ is weakly dependent on the frequency. Going therefore from right to left in the horizontal axis of Fig.\ref{fig:theta_i_module}, sooner or later $\omega_d$ reaches values lower than  $|\gamma_{RB,d}|$ so that the second side of \eqref{eq:theta_i} becomes almost independent from $f_d$. 
The same considerations apply for the other curves $|M_d|=constant$, with the difference that the values of the frequencies at which  the lines become independent from $f_d$ decrease for increasing values of $|M_d|$ (or increase for decreasing values of $|M_d|$). This behavior can be intuitively explained, since an increase of $|M_d|$ determines an increase of $X_{d,k}$ (see \eqref{eq:X_i_k}) which in turn brings to a reduction of the terms  $\frac{2J_1(X_{d,k})}{X_{d,k}}$ of the sum  present in \eqref{eq:gamma_RB}, i.e. a decrease of the ``cutoff angular frequency'' $|\gamma_{RB,d}|$.  

Still in Fig. \ref{fig:theta_i_module} the behavior of $|M_d|$ in case of absence of RB is also reported, which puts into evidence the described effect of the RB field re-injection, namely the impossibility at a given low frequency to obtain high values of $|M_d|$ for dithering current amplitudes lower than a minimum ``threshold'' value.

\begin{figure}[t]
\centering
\includegraphics[scale=0.33]{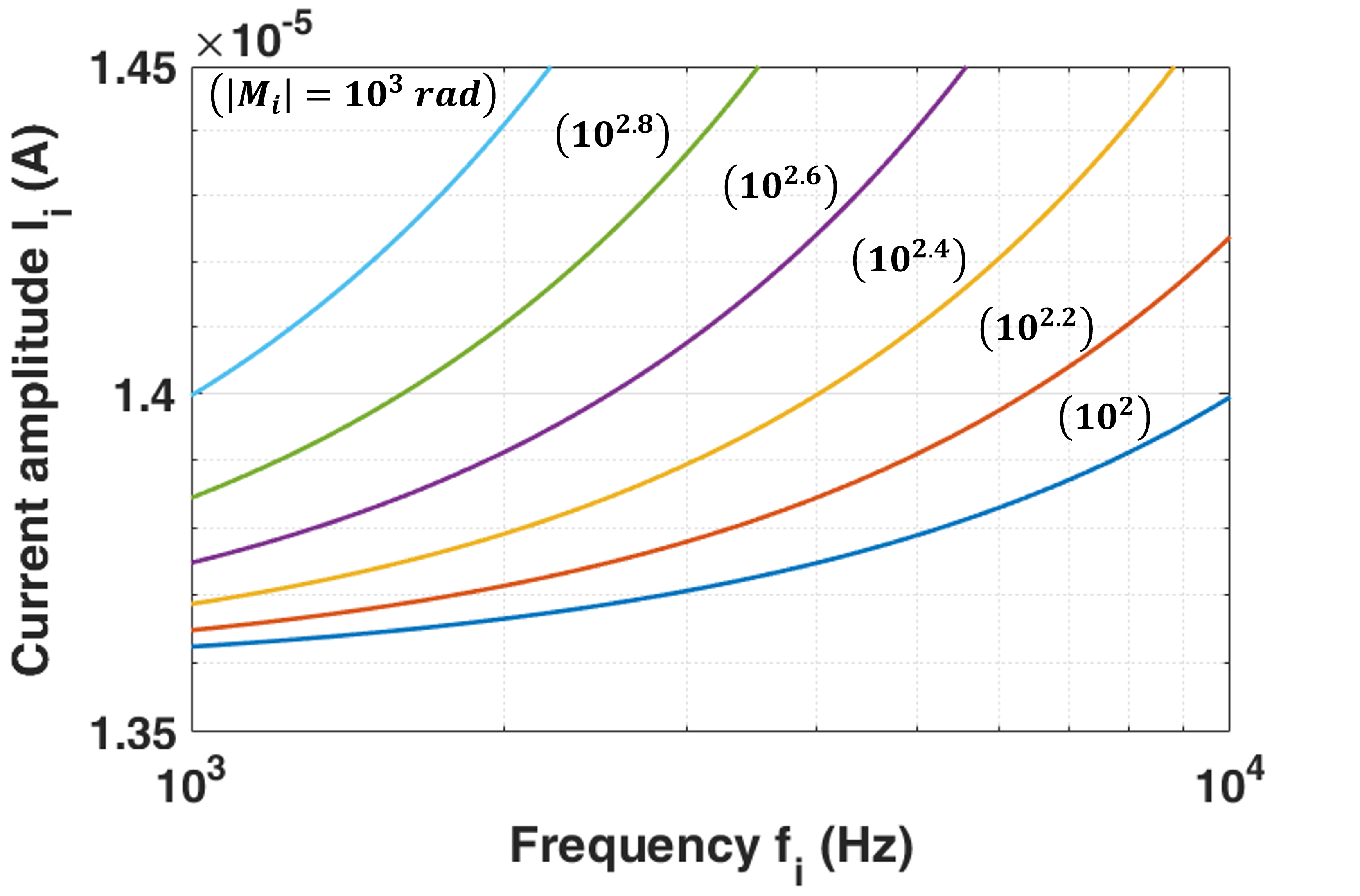}
\caption{Values of $|M_i|$ which are reached in the interval $f_i \in [1, 10]$ kHz in the vicinity of the values of $I_i$ where the curves of Fig.\ref{fig:theta_i_module} appear to exhibit an almost horizontal behavior. Note that the subscripts can be assumed as $i=RF$ or $i=d$.}
\label{fig:zoom_theta_i_module}
\end{figure}

Note however that minimal increases of $I_d$ from this ``threshold'' value lead to very high variations of $|M_d|$. This can be appreciated in Fig. \ref{fig:zoom_theta_i_module}, which focalizes the same quantities represented in  Fig. \ref{fig:theta_i_module} in limited ranges of $f_d$ and $I_d$.

Given the mentioned requirements for an effective application of the {\it dithering tone} technique (high $|M_d|$, $I_d$, $f_d$ as low as possible), the practical indication is then to utilize a value of $I_d$ around such "threshold" value, while keeping $f_d$ as low as possible. 
A further physical phenomenon will be however highlighted in the next Section, which prevents the choice of arbitrarily low values for $f_d$.

\subsection{Power Spectral Density at the Receiving End and effects of the {\it Dithering Tone} characteristics} 
\label{sub:final_spectrum}
As mentioned at the end of Section II.A, at the final end $z=L$ of the fiber connection, the field received by the PIN photodetector is given by the sum $E_{TX}(t,L)+E_{RB2}(t,L)$, with the first addend given by \eqref{eq:fiber_field} with $\tau=\tau_L=L/v_g$, and the second one given by:
\begin{eqnarray}
&&E_{RB2}(t,L) =\frac{\sqrt{R_{l,2}(1-R_{l,2})}\cdot \eta_c^{3/2}}{\sqrt{\alpha_{iso}}} \times \nonumber\\
&&\times \sum_{k=1}^K \rho _k {{E}}(t-2 \tau _k-\tau_L)e^{-\alpha v_g (2 \tau _k+\tau_L)}
\label{eq:E_RB2_L}
\end{eqnarray}


The current generated by the photo-detector can then be computed as:
\begin{eqnarray}
&i_{out}(t)&=\mathcal{R}\left|E_{TX}(t,L) + E_{RB2}(t,L)\right|^2\nonumber\\
&&=\mathcal{R}\cdot \left|E_{TX}(t,L)\right|^2+\mathcal{R}\cdot\left|E_{RB}(t,L)\right|^2+\nonumber\\
&&+\mathcal{R}\cdot 2\Re  {\it \bf e}  \{E_{TX}(t,L)\cdot E_{RB}^*(t,L)\}
\label{eq:i_out_00}
\end{eqnarray}
where $\mathcal{R}$ is the Responsivity of the detector.

At the right-end-side of Equation \eqref{eq:i_out_00}, the first addend would coincide with the total detected current if the RB effect were absent. Moreover, the term $|E_{RB}(t,L)|^2$ is much smaller with respect to the others therefore the second added will be neglected in the following. 

In order to analyze the RB-induced spurious terms in the current power spectrum, only the last addend at the right hand side of \eqref{eq:i_out_00} has then to be considered, which will be named as $i_{out_{TX,RB}}$. From Equations \eqref{eq:fiber_field} (with $\tau=\tau_L$) and  \eqref{eq:E_RB2_L},  exploiting the prostaferesis formula $\cos(u)-\cos(v)=-2\sin\left( \frac{u-v}{2}\right) \sin\left( \frac{u+v}{2} \right)$ with $u=\omega_i(t-\tau_L)+\angle M_i$ and $v=\omega_i(t-\tau_L-2\tau_k)+\angle M_i$, its expression results to be:
{\begin{align}
&i_{out_{TX,RB}}(t)\simeq \mathcal{R}\sqrt{\frac{R_{l,2}}{\alpha_{iso}}}(1-R_{l,2})\eta_c^2|E(t-\tau_L)|  \times \label{eq:i_out_TX_RB_no_approx}\\
&\times 2\Re  {\it \bf e}  \left\{ 
\sum_{k=1}^{k_{max}}\rho^*_k |E(t- 2\tau_k -\tau_L)| e^{-\alpha v_g 2\tau_k+j\omega_{opt}2\tau_k}  \times \right.\nonumber\\
&\times \left.e^{j\left\{-\sum_i  X_{i,k}sin\left[\omega_i(t-\tau_k-\tau_L)+\angle M_i\right] + \phi(t-\tau_L)-\phi(t-2\tau_k-\tau_L)\right\}}\right\} \nonumber
\end{align}}
where $X_{i,k}$ has been defined in \eqref{eq:X_i_k}.

\begin{figure}[t]
\centering
\includegraphics[scale=0.42]{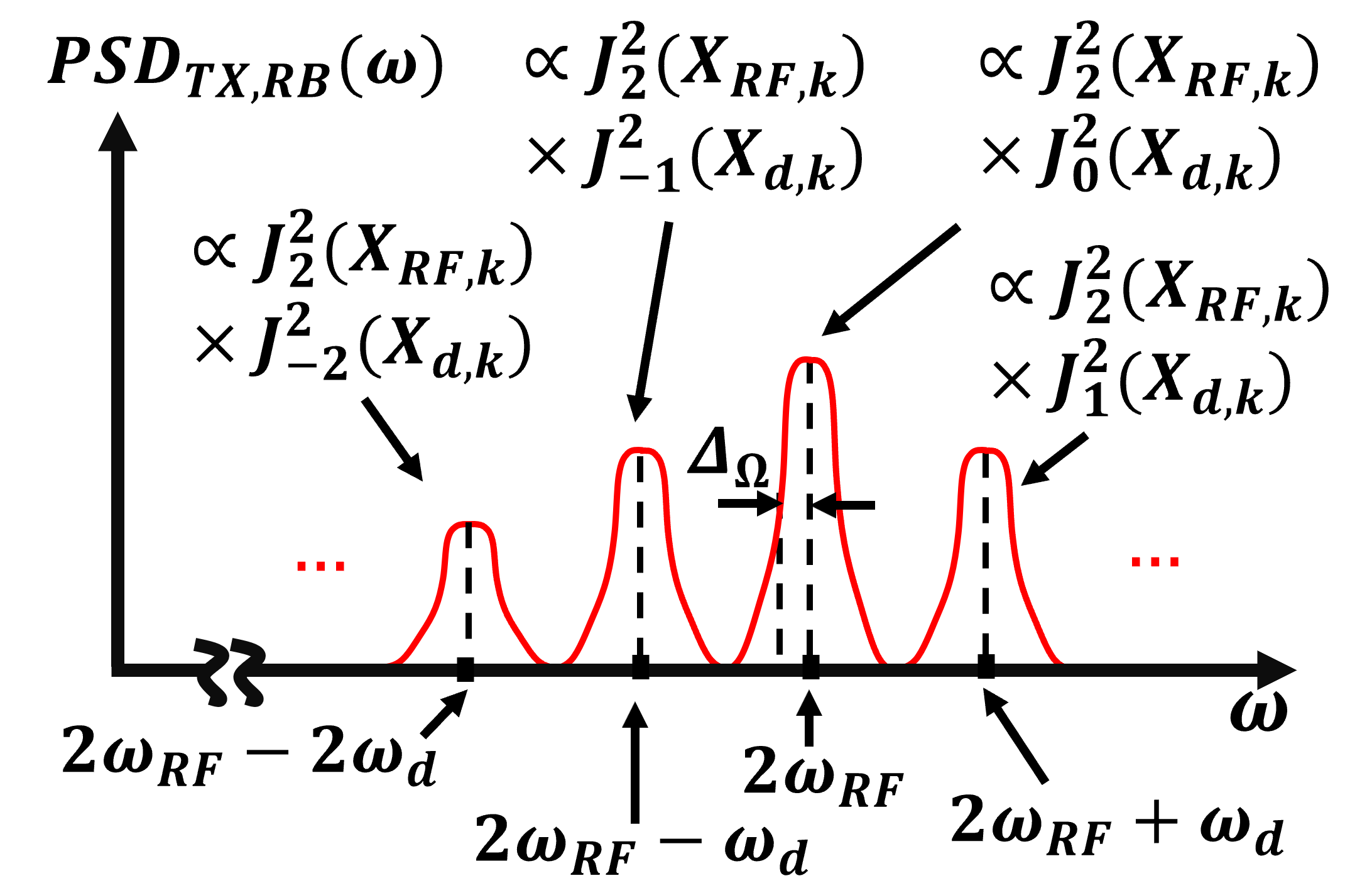}
\caption{Qualitative representation of the behavior of $PSD_{TX,RB} (\omega)$ in the vicinity of $\omega= 2 \omega_{RF}$, represented by \eqref{eq:PSD_TX_RB} with $n_{RF}=2$, for the $k-th$ elementary RB reflection, when $\omega_{d}$ is sufficiently greater than  $\Delta_{\Omega}$. In this case the power of the second-harmonic distortion term $P_{2\omega_{RF}}$ practically consists in the only element corresponding to $n_{d}=0$.}
\label{fig:PSD_1}
\end{figure}

In line with the perturbative approach adopted, in \eqref{eq:i_out_TX_RB_no_approx} both field modules can be assumed to be equal to $E_0$. Collecting all the constants in a unique term $\Upsilon=\mathcal{R}\sqrt{\frac{R_{l,2}}{\alpha_{iso}}}(1-R_{l,2})\eta_c^2E_0^2$, and exploiting the Jacobi-Anger expansion $e^{-ju\sin(v)}=\sum_{n=-\infty}^{\infty}J_{n}(u)e^{-jnv}$, 
where $u=X_{i,k}$ and $v=\omega_{i}(t-\tau_L-\tau_k) + \angle M_i$, taking advantage of \eqref{eq:phase_noise} and \eqref{eq:rho_k}, after a lengthy but straightforward derivation,  the expected value of the power spectral density $PSD_{TX,RB}$ of $i_{out_{TX,RB}}(t)$ assumes the following expression: 

\begin{align}\label{eq:PSD_TX_RB}
& PSD_{TX,RB} (\omega)=  \nonumber \\
&={F_{\xi}} \left \{ \mathbb{E} \left[ \left \langle i_{out_{TX,RB}}(t)i_{out_{TX,RB}}^{\ast}(t-\xi) \right \rangle \right] \right\}
\simeq \nonumber \\ 
&\simeq \Upsilon^2 \sum_{k=1}^{k_{max}}{\sigma^2_{\rho_k}} e^{-4\alpha v_g \tau_k}
\sum_{\substack{n_{d},n_{RF}=\\=-\infty}}^{+\infty} {J_{n_{d}}^2(X_{d,k}) J_{n_{RF}}^2(X_{RF,k})} \times
 \nonumber \\ 
&\times \frac{2\Delta_{\Omega}}{\Delta_{\Omega}^2+[\omega-(n_{d}\omega_{d}+n_{RF}\omega_{RF})]^2}
\end{align}

Focusing for example in the vicinity of $\omega=2\omega_{RF}$, i.e. considering $n_{RF}=2$ in \eqref{eq:PSD_TX_RB}, and referring to the generic contribution related to the $k-th$ elementary RB reflection, the function $PSD_{TX,RB} (\omega)$,  as can be appreciated in Fig. \ref{fig:PSD_1},  consists in the superposition of lorentzian functions centered in $2\omega_{RF}, 2\omega_{RF} \pm \omega_{d}, 2\omega_{RF} \pm 2\omega_{d}, \dots$ with amplitudes proportional respectively to  $J_2^2(X_{RF,k})J_0^2(X_{d,k})$,  $J_2^2(X_{RF,k})J_{\pm 1}^2(X_{d,k})$, $J_2^2(X_{RF,k})J_{\pm 2}^2(X_{d,k}), ...$. 

The power of the undesired second-harmonic distortion term $P_{2\omega_{RF}}$  within an elementary bandwidth $\delta_{\Omega}$ centered in $2 \omega_{RF}$ can be directly evaluated as the integral of the Power Spectral Density multiplied by the load resistance $R_L$: $P_{2\omega_{RF}}=R_L\cdot\int_{2\omega_{RF}-\delta_{\Omega}/2}^{2\omega_{RF}+\delta_{\Omega}/2} PSD_{TX,RB} (\omega) d\omega$ and results to be given by the last side of \eqref{eq:PSD_TX_RB} multiplied by $\delta_{\Omega}$. 

Note that if the {\it dithering tone} were absent, the contribution to $P_{2\omega_{RF}}$ coming from each elementary RB reflection would be just proportional to $J_2^2(X_{RF,k})$ and its impact could be reduced  increasing $|M_{RF}|$, i.e. appropriately operating on $\omega_{RF}$ and $I_{RF}$, operations which would all result as not realizable, e.g. in the case where the modulating signal consists in an uncontrollable RFI disturbance.
 
On the contrary, in presence of the {\it dithering tone} (see again the case represented in Fig. \ref{fig:PSD_1}) the contribution to $P_{2\omega_{RF}}$ coming from each elementary RB reflection is proportional to $J_2^2(X_{RF,k}) J_0^2(X_{d,k})$. In this case, it is possible to reduce its impact reducing the value of $J_0^2(X_{d,k})$, i.e. increasing $|M_{d}|$, and the previous Section II-B has been just devoted to determine how to obtain high values of $|M_d|$ while keeping $I_d$ and $f_d$ (or equivalently $\omega_d$) at their lowest possible levels.

\begin{figure}[t]
\centering
\includegraphics[scale=0.42]{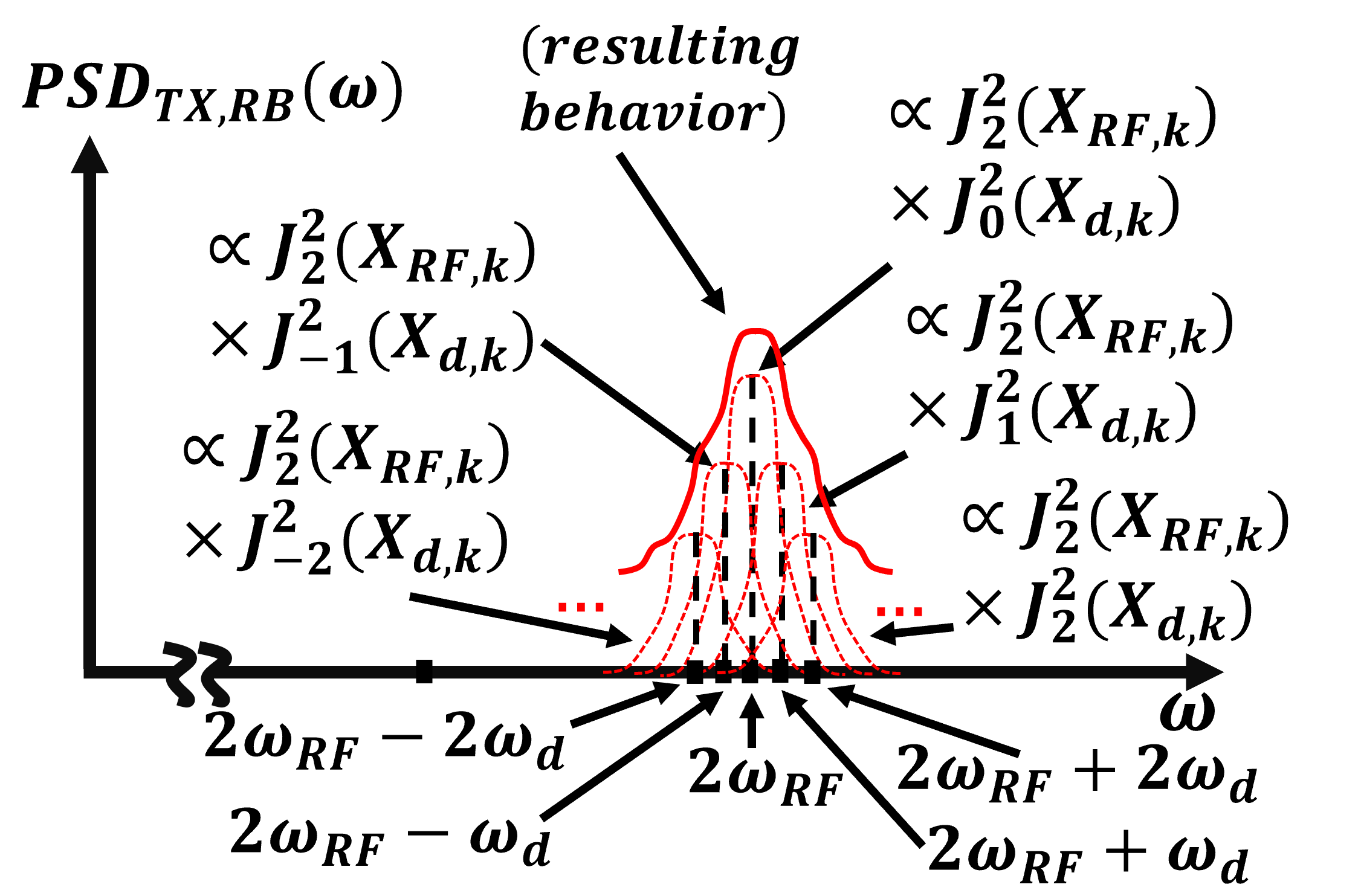}
\caption{Same as Fig.\ref{fig:PSD_1} with the difference that $\omega_{d}$ has been so decreased that the ``tails'' of the elements corresponding to $n_{d} \not = 0$ of the sum in $n_{d}$ of \eqref{eq:PSD_TX_RB} contribute to the term at frequency $2f_{RF}$.}
\label{fig:PSD_2}
\end{figure}

\begin{figure}[t]
\centering
\includegraphics[scale=0.43]{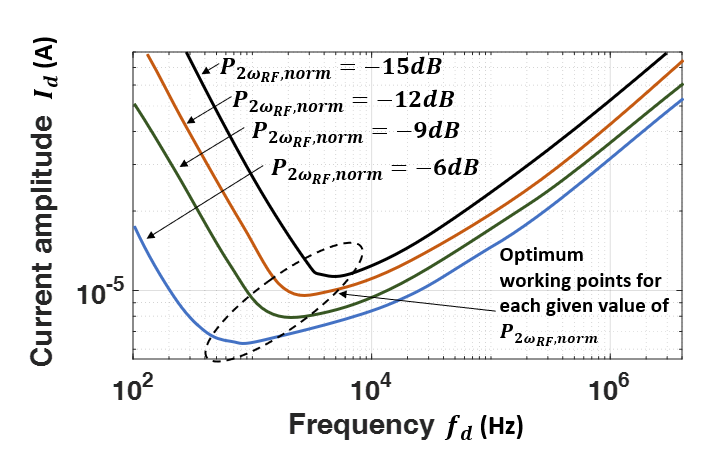}
\caption{Simulated behavior of the values of current $I_i$ and frequency $f_i$ which gives the normalized second-harmonic power $P_{2\omega_{RF},norm}=cost$. Both current and frequency axis are plotted in log-scale for a better view of the whole behavior.}
\label{fig:final_comparison}
\end{figure}

However, continuing to decrease $\omega_d$, the ``tails'' of the higher order terms of the sum in $n_{d}$ of \eqref{eq:PSD_TX_RB}  (i.e. those corresponding  to $n_{d}= \pm 1, \pm 2, \dots $) result to contribute to the term at frequency $2 \omega_{RF}$ and consequently prevent the reduction of the power at $2\omega_{RF}$ which is pursued with the {\it dithering tone} introduction. The phenomenon is qualitatively represented in Fig. \ref{fig:PSD_2} and its effects are illustrated in Fig. \ref{fig:final_comparison}, which shows the simulated behavior of the values of $P_{2\omega_{RF}}$ normalized to its maximum, named in the figure $P_{2\omega_{RF},norm}$, covering a wide range of dithering frequencies, i.e, $f_d \in[100-4\times10^6]$ Hz.

The value of $\omega_d$ cannot be reduced at will, for the insurgence of this additional phenomenon and an optimum working point can be identified for the application of the \emph{dithering tone} technique through which the minimum value of $I_d$ can be applied t guarantee a given low level of $P_{2\omega_{RF}}$, while keeping $f_d$ as low as possible.

The considerations just developed with reference to the second harmonic term can be straightforwardly applied to the other spurious frequency terms (e.g. at frequencies $3f_{RF}$,$f_{RF}+f_{d}$, etc...).

\section{Experimental Measurements and discussion}\label{sec:simul_meas}
\begin{figure}[hbtp]
\centering
\includegraphics[scale=0.4]{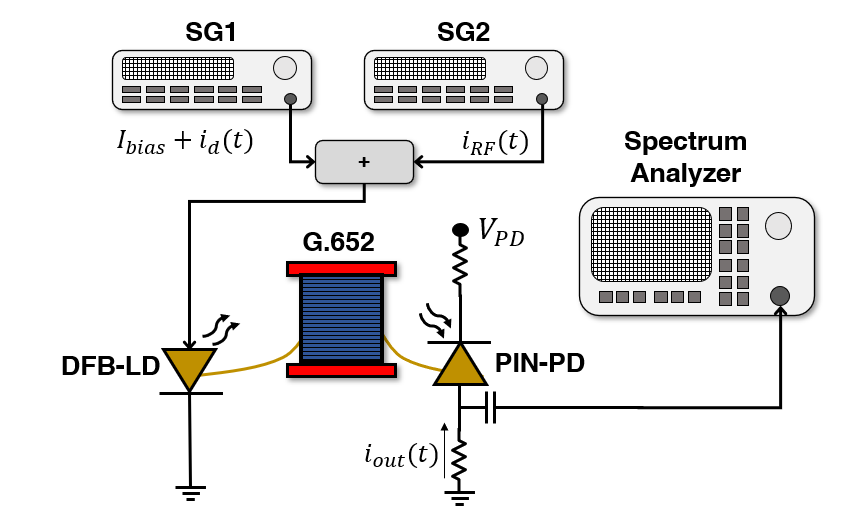}
\caption{Experimental setup utilized. The setup is composed of two Signal Generators (SG1 and SG2), an RF passive coupler, a Distributed FeedBack Laser Diode (DFB-LD), a span of G.652 fiber, a Positive-Intrinsic-Negative PhotoDiode (PIN-PD) and a Spectrum Analyzer.}\label{fig:setup}
\end{figure}

With the aim of confirm the theoretical results derived in previous section, measurements have been performed using the setup illustrated in Figure \ref{fig:setup}. 
\begin{table}[hbpt]
\centering
\caption{Parameters of the link under test.}
\renewcommand\arraystretch{1.2}{
\begin{tabular}{c|c|c}
\textbf{Symbol}& \textbf{Physical meaning}& \textbf{Value}\\
\hline
$I_{th}$        & Laser threshold current              & 9 $mA$\\
$I_{bias}$      & Laser bias current                   & 37 $mA$\\
$P_{opt}$       & Laser output power                   & $6\,dB_m$\\
$\alpha_{iso}$  & Laser power isolation                & 20 $dB$\\
$\Delta_{\Omega}$& Laser Coherence angular frequency band & $3.14\cdot 10^6 \frac{rad}{s}$\\
$\alpha$        & Fiber field attenuation factor       & $5^{-5}$ $\frac{neper}{m}$\\
$\eta_c$		& Power coupling lens coefficient	&0.4\\
$\mathcal{S}$   & Fiber Recapture factor               & $10^{-3}$\\
$\mathcal{R}$   & PD Responsivity                & 1 $\frac{A}{W}$\\
$R_L$           & Load Resistance                & 50 $\Omega$\\
\hline
\end{tabular}}\label{tab:parameters}
\end{table}

The RoF link under test is composed of a DFB laser working at 1310nm connected to a span of G.652 fiber which is directly connected to a PIN Photodiode biased at $V_{PD}=5$ Volts. The characteristics of laser and photodiode are reported in Table \ref{tab:parameters}. Two separate signal generators (SGs) are used to supply the laser. In particular, SG1 is used to supply the bias current $I_{bias}$ and the dithering tone $i_{d}(t)$ where both amplitude $I_{d}$ and frequency $f_{d}$ have been varied. The SG2 instead, is used to generate the RF tone $i_{RF} (t)$ at the frequency $f_{RF}=70$ MHz, simulating then a possible RFI falling in an interval of interest for low-frequency radioastronomic facilities. The three currents are then coupled using a passive coupler which is connected directly to the Laser. Indeed, in order to set the desired amplitudes of the three currents, the insertion loss of the coupler has been taken into account so that in the following discussion the values of $I_{bias}$ and the amplitudes of $i_{d}(t),i_{RF} (t)$ are considered ones injected into the laser, i.e. after the coupler.
After propagating in a standard ITU-T G.652 optical fiber of length $L=5$ km, which is in the order of the maximum distances covered in SKAlow \cite{SPIE}, the signal is detected by photodiode and through a spectrum analyzer the amplitude levels of the spurious tones generated at frequencies $2f_{RF}$ and $f_{RF}+f_{d}$ are evaluated.

\begin{figure}[htbp]
\centering
\includegraphics[scale=0.38]{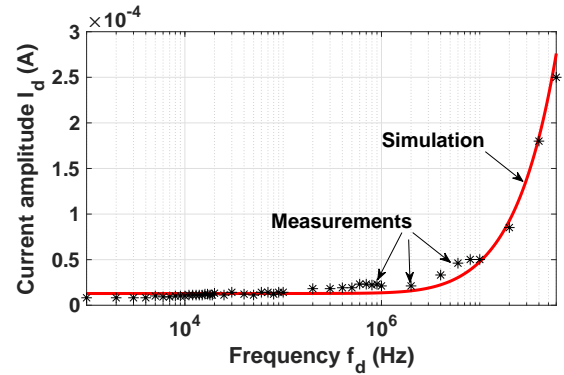}
\caption{Comparison between experimental and simulated values of the modulating current amplitude $I_d$ which, for a given modulating frequency $f_d$, determine the fixed value of the phase modulation index $|M_d|=0.92$.}
\label{Fig:confronti}
\end{figure}

\begin{figure}[htbp]
\begin{subfigure}{.5\textwidth}
  \centering
  \includegraphics[width=\linewidth]{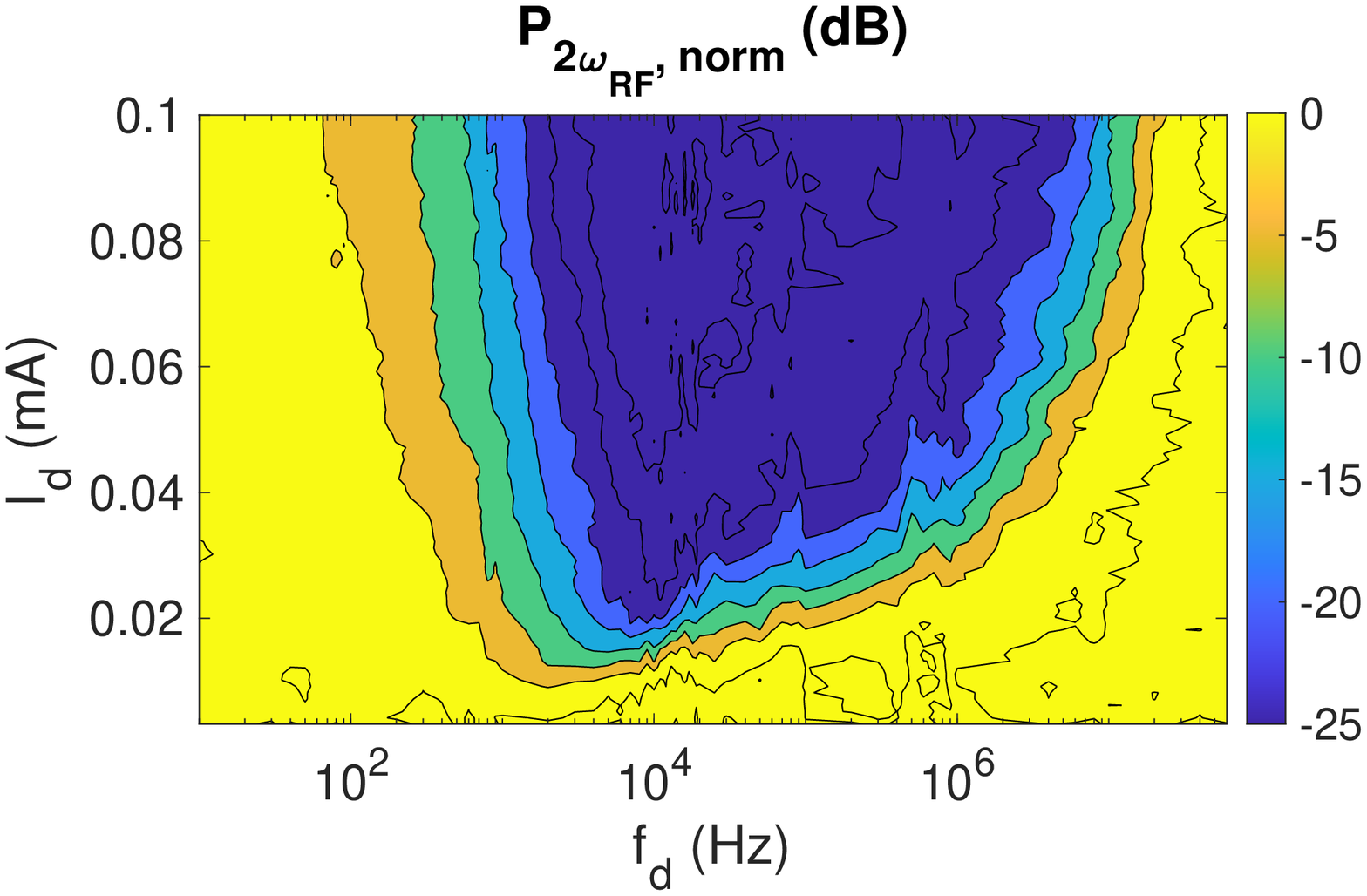}  
  \caption{}
  \label{fig:Meas_HD2}
\end{subfigure}
\begin{subfigure}{.5\textwidth}
  \centering
  \includegraphics[width=\linewidth]{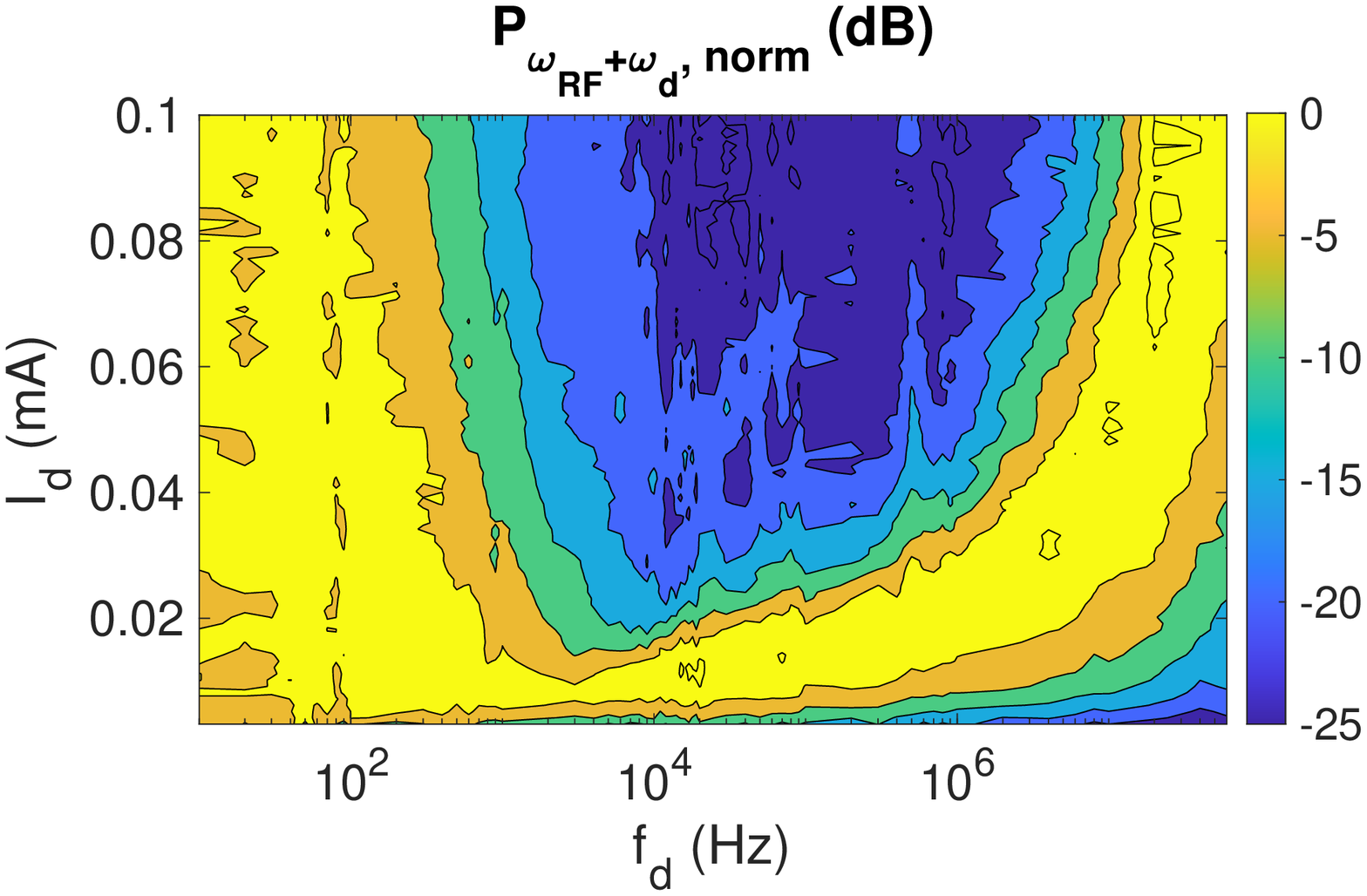}  
  \caption{}
  \label{fig:Meas_IMD2}
\end{subfigure}
\caption{Contour map of the measurement of the power level at the frequency $2\omega_{RF}$ (a) and $\omega_{RFI} + \omega_{d}$ (b) employing 5Km of G.652 fiber length as a function of $I_{d}$ and $f_{d}$. The level of the optimum dithering frequency can be found in this case for about 10 kHz.}
\end{figure}

\begin{figure}[htbp]
\begin{subfigure}{.5\textwidth}
  \centering
  \includegraphics[width=\linewidth]{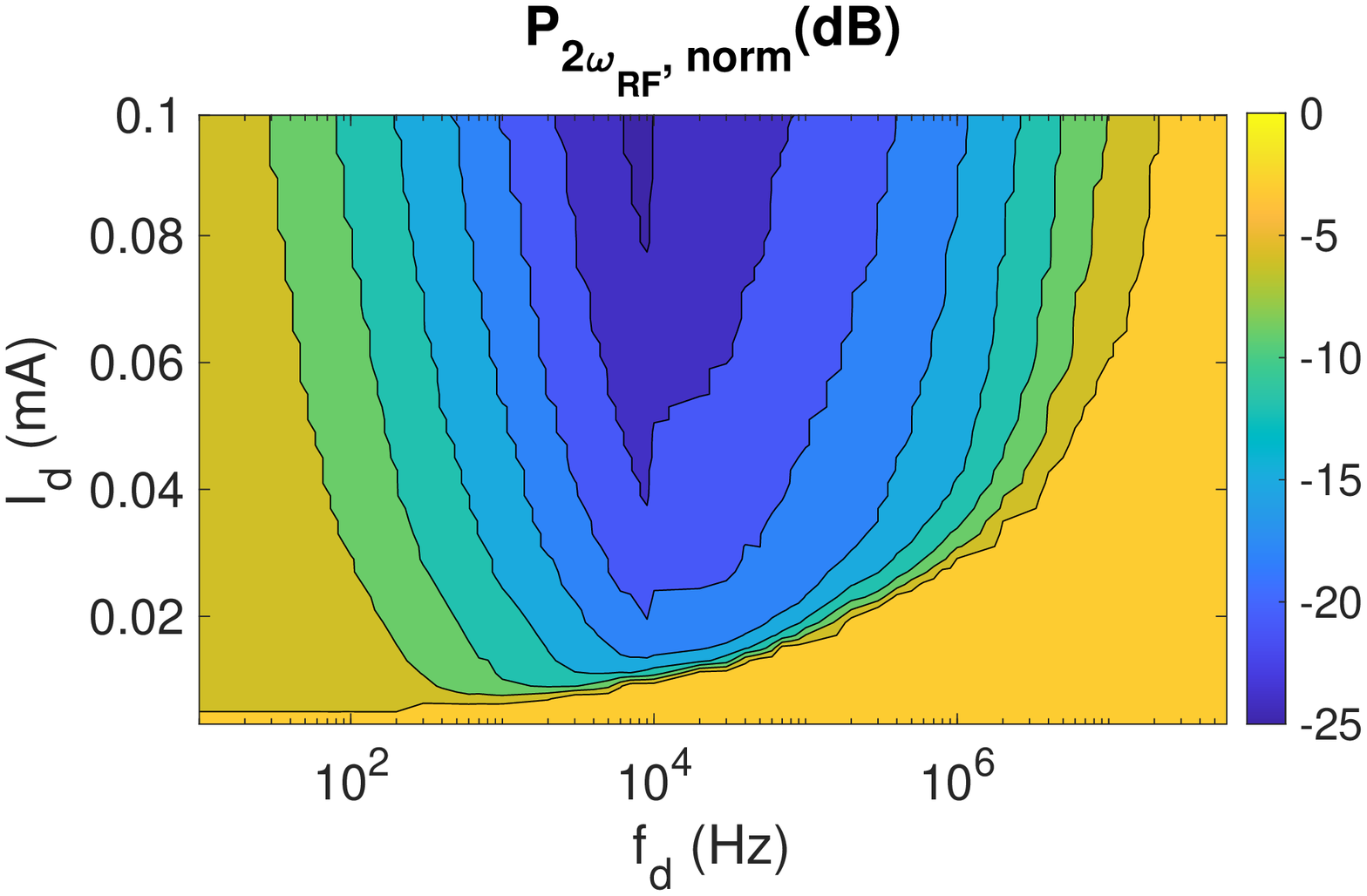}  
  \caption{}
  \label{fig:sim_HD2}
\end{subfigure}
\begin{subfigure}{.5\textwidth}
  \centering
  \includegraphics[width=\linewidth]{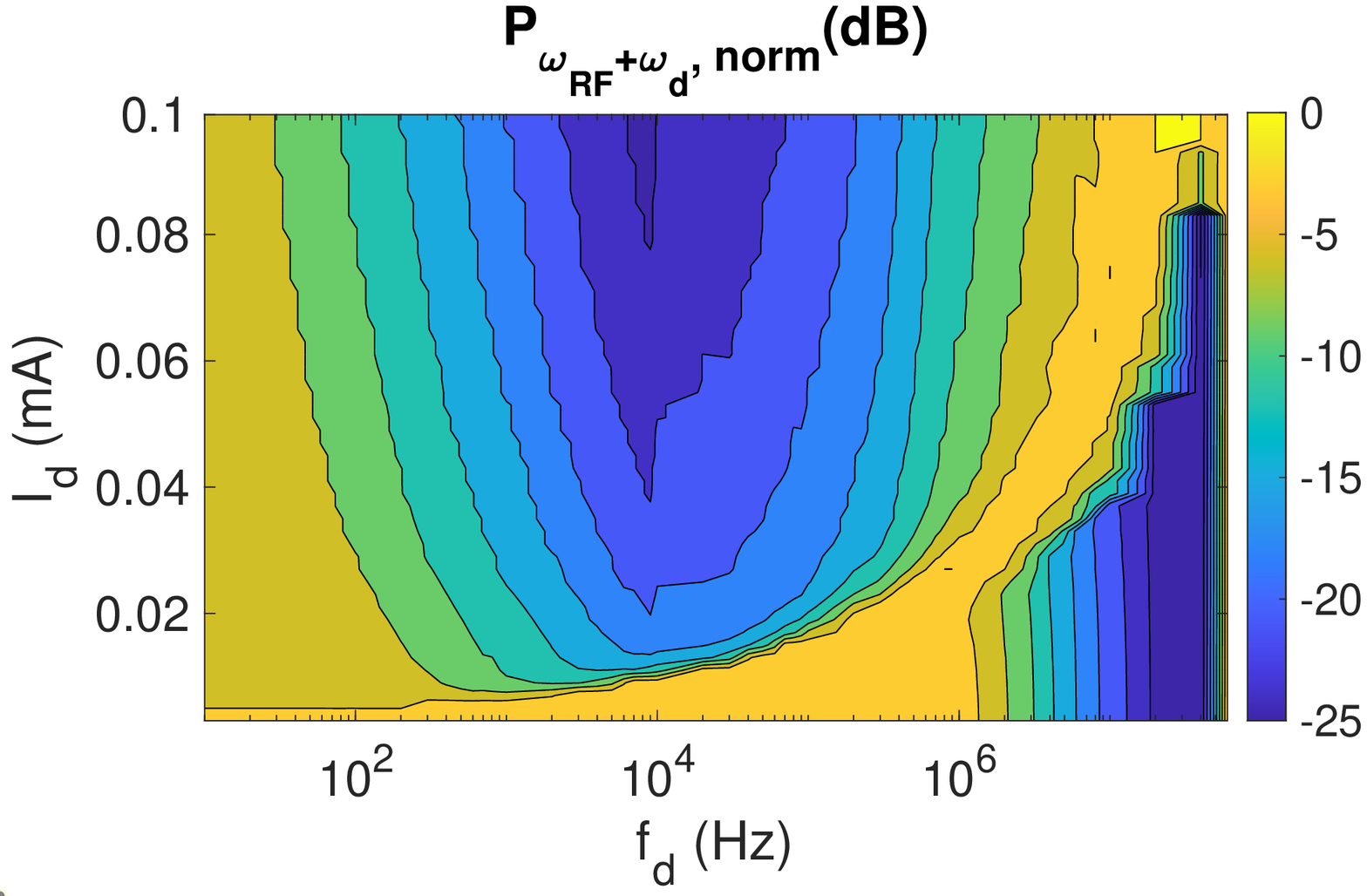}  
  \caption{}
\label{fig:sim_IMD2}
\end{subfigure}
\caption{Contour map of the simulation of the power level at the frequency $2\omega_{RF}$ (a) and $\omega_{RF} + \omega_{d}$ (b) employing 5 km of G.652 fiber length as a function of $I_{d}$ and $f_{d}$. The level of the optimum dithering frequency can be found in this case for about 10 kHz.}
\end{figure}

As a preliminary test, the values of the dithering current amplitude $I_{d}$ which lead to a maximum of the intermodulation term at frequency $\omega_{RF}+\omega_{d}$, $P_{\omega_{RF}+\omega_{d}}$ have been measured for varying values of $f_d$. The behavior of this spurious component is described by \eqref{eq:PSD_TX_RB} considering $n_d=1,n_{RF}=1$ and it can be shown that the value of current $I_d$ which gives the maximum of  $P_{\omega_{RF}+\omega_{d}}$ is obtained when $M_d\simeq 0.92$, which, with good approximation, corresponds to the value which gives the maximum of $\sum_{k=1}^{k_{max}}J_1^2(X_{d,k})$. Figure \ref{Fig:confronti} shows the good agreement between simulated and measured values of $I_d$.

Note that this analysis have been performed considering $f_d=1$ kHz as the lowest value of dithering frequency. The reason behind this choice is that for lower frequencies it is not possible to identify experimentally the maximum of $P_{\omega_{RF}+\omega_{d}}$. 

After this preliminary test, an analysis of the behavior of the terms $P_{2\omega_{RF}}$ and $P_{\omega_{RF}+\omega_{d}}$ as a function of  $I_d$ and $f_d$ has been experimentally performed. 

The measurements have been conducted by fixing the value of certain frequency $f_d$ and sweeping the current $I_d$ for a range $[0.001-0.1]$ mA. This sweeping has been then repeated for each frequency $f_d$ of interest, generating in this way the maps shown in Figures \ref{fig:Meas_HD2} and \ref{fig:Meas_IMD2}. Those maps represent the normalized values of $P_{2\omega_{RF}}$ and $P_{\omega_{RF}+\omega_{d}}$, named in the figures $P_{2\omega_{RF}, norm}$ and $P_{\omega_{RF}+\omega_{d}, norm}$ respectively, measured at the output of the photodetector through the spectrum analyzer (see again Fig.\ref{fig:setup}).
An analogous map has been performed also by implementing the model shown in previous section, and the results are represented in Fig. \ref{fig:sim_HD2} and \ref{fig:sim_IMD2}. The maps show the different levels of harmonic and intermodulation distortion which offer a good reproduction of the experimental results, confirming the correctness of the theoretical study performed. 
In particular, both simulations and experimental results show that for the considered link, a frequency $f_{d}\simeq 10$ kHz and an amplitude $I_{d} \simeq 0.04$ mA constitute the optimum working point which allows to reduce the powers of both second harmonic distortion term $P_{2\omega_{RF}}$ and second order intermodulation product $P_{\omega_{RF}+\omega_d}$ by at least 25 dB referred to their respective maximum values.


%

\section{Conclusion}
An exhaustive study of the detrimental nonlinear effects produced by the Rayleigh Backscattering present within Analog Radio-over-Fiber link, and of the technique used to mitigate them, namely the \emph{dithering technique}, has been shown. As a first important and novel aspect, a detailed description of the behavior of a directly modulated laser in the presence of optical feedback due to Rayleigh Backscattering has been presented, showing the significant changes produced on the laser spurious phase modulation (i.e. frequency chirp) characteristics and on the power spectrum of the signal received at the end of the fiber connection. Then, it has been shown that these changes impact significantly on the optimum characteristics of current and frequency of the sinusoidal dithering tone. The technique showed to be particularly effective on the typical scenarios of low frequency radioastronomic systems (for example, it has been applied in the demonstrators of SKA-low), which suffers to a relatively great degree the impact of RB due to their peculiar characteristics. 

\appendix
\subsection{Derivation of $\Gamma_S, \Gamma_{\theta}$}
\label{app:1}
The rate equation of the complex electrical field $E(t)$ in presence of RBS can be written as follows \cite{Chraplyvy}:

\begin{align}
&\frac{dE(t)}{dt}=\label{eq:Kobayashi_Rayleigh}\\
&=\left[j\omega(N) +  \frac{1}{2}\left(G(N,S) -\frac{1}{\tau_p}\right)\right]E(t)+\sum_{k=1}^{K}C_kE(t-2\tau_k)\nonumber
\end{align}

where the last term on the right-hand-side represents the RBS contribution. Putting $E(t)=|E(t)|e^{j[\omega_0t+ \theta(t)]}$ and creating two equations, one for the real part and one for the imaginary part it results:
\begin{eqnarray}
&& |E(t)|\frac{d|E(t)|}{dt}= \frac{1}{2}\left(G(N,S) -\frac{1}{\tau_p}\right)|E(t)|^2+ \nonumber \\
&& = +\sum_{k=1}^K|C_k| |E(t)||E(t-2\tau_k)|\cdot\nonumber\\
&&\cdot\cos[\omega_{0}\cdot 2\tau_k +\theta(t)-\theta(t-2\tau_k)-\angle \rho_k]
\label{eq:app1_1}
\end{eqnarray}
and:
\begin{eqnarray}
&& \frac{d\theta(t)}{dt}= \left[\omega(N) -\omega_0 \right]-\sum_{k=1}^K|C_k| \frac{|E(t-2\tau_k)|}{|E(t)|}\cdot\nonumber\\
&&\cdot\sin[\omega_{0}\cdot 2\tau_k +\theta(t)-\theta(t-2\tau_k)-\angle \rho_k]
\label{eq:app1_2}
\end{eqnarray}
Exploiting the relationship $S(\cdots)=|E(\cdots)|^2$, $|E(t)|\frac{d|E(\cdots)|}{dt}=\frac{1}{2}\frac{dS(\cdots)}{dt}$ the expressions given by \eqref{eq:Gamma_S}, \eqref{eq:Gamma_PHI} can be obtained

\subsection{Derivation of $\gamma_{S,i}, \gamma_{\theta,i}$}
\label{app:2}
Starting from \eqref{eq:Gamma_S} and exploiting \eqref{eq:delta S}, \eqref{eq:delta theta}, it is:
\begin{eqnarray}
&&\Gamma_S(t)=2\sum_{k=1}^K|C_k| \left( \left[S_0+\Delta S +S_i\cos(\omega_i t +\angle S_i)\right]\cdot \right.
\nonumber \\
&& \left. \cdot\left[S_0+\Delta S +|S_i| \cos(\omega_i (t-2\tau_k) +\angle S_i)\right]\right)^{\frac{1}{2}}
\cdot\nonumber\\
&&\cdot\cos\left[(\omega_{0}+\Delta \omega)\cdot 2\tau_k -\angle \rho_k + \right.
\nonumber \\
&&\left. -|M_i|\cdot 2\sin(\omega_i 2\tau_k)\sin\left(\omega_i(t-2\tau_k)+\angle M_i\right)\right]
\label{eq:gamma_Si_1}
\end{eqnarray}

From assumption \eqref{eq:perturb_1} the terms $\Delta S$ and $S_i$ can be neglected under the square root. Putting then $\psi_k=(\omega_{0}+\Delta \omega)\cdot 2\tau_k -\angle \rho_k$  and  exploiting the truncated Jacobi-Anger expansion $e^{-ju\sin(v)}= J_0(u) + 2J_1(u)\sin(v)$ , with $u=2|M_i|\sin(\omega_i \tau_k)$ and $v=\sin\left(\omega_i(t-\tau_k)+\angle M_i\right)$, in order to focus on the terms at $DC$ and at $\omega_i$, it results:
\begin{eqnarray}
&&\Gamma_S(t)=2 S_0\sum_{k=1}^K|C_k| \cos(\psi_k)\cdot J_0\left[2|M_i|\sin(\omega_i \tau_k)\right]+ \nonumber \\
&& + 2 S_0 \sum_{k=1}^K|C_k|\sin(\psi_k)\cdot \nonumber \\
&&\cdot 2J_1\left[2|M_i|\sin(\omega_i \tau_k)\right]
\cdot \sin\left[\omega_i(t-\tau_k)+\angle M_i\right]
\label{eq:gamma_Si_2}
\end{eqnarray}
Putting now $X_i=2|M_i|sin(\omega_i \tau_k)$ and elaborating further the term written in the last line of \eqref{eq:gamma_Si_2} it results:
\begin{eqnarray}
&& 2J_1\left[2|M_i|\sin(\omega_i \tau_k)\right]
\cdot \sin\left[\omega_i(t-\tau_k)+\angle M_i\right]=
\nonumber \\
&&= \frac{2 J_1(X_i)}{X_i}
\cdot 2|M_i|\sin(\omega_i \tau_k) sin\left[\omega_i(t-\tau_k)+\angle M_i\right]
 \nonumber \\
 &&= \Re {\it \bf e} \left\{ 2\frac{J_1(X_i)}{X_i}
\cdot 2|M_i| \frac{e^{j\omega_i\tau_k}-e^{-j\omega_i\tau_k}}{2j} \right.\cdot
 \nonumber \\
 && \left. \cdot (-j)e^{j\left[\omega_i(t-\tau_k)+\angle M_i\right]}\right\}
 \nonumber \\
 &&= \Re {\it \bf e} \left\{ -2\frac{J_1(X_i)}{X_i}
\cdot \left[1-e^{-j2\omega_i\tau_k} \right] \right.\cdot
 \nonumber \\
 && \left. \cdot M_ie^{j\left[\omega_i t\right]}\right\}
\label{eq:gamma_Si_3}
\end{eqnarray}
Therefore, the second addend of $\Gamma_S(t)$ reported in the two last lines of \eqref{eq:gamma_Si_2}, which contributes to the system at angular frequency $\omega_i$ in \eqref{eq:system} becomes:
\begin{eqnarray}
&& - 2 S_0\sum_{k=1}^K|C_k| \sin(\psi_k)\cdot \frac{2 J_1(X_i)}{X_i}
\cdot \left[1-e^{-j2\omega_i\tau_k} \right] \cdot \nonumber \\
&&  \cdot M_ie^{j\omega_i t}= -2 S_0 \cdot \gamma_{S,i} \cdot  M_ie^{j\omega_i t}
\label{eq:gamma_Si_4}
\end{eqnarray}
from which the expression of $\gamma_{S,i}$ can be recognized.

Proceeding in a  similar fashion with $\Gamma_{\theta}$, starting from \eqref{eq:Gamma_PHI} and exploiting \eqref{eq:delta S}, \eqref{eq:delta theta} and the condition \eqref{eq:perturb_1}, it is:
\begin{eqnarray}
&& \Gamma_{\theta}= -\sum_{k=1}^K|C_k| \cos(\psi_k)\cdot
2J_1\left[2|M_i|\sin(\omega_i \tau_k)\right] \cdot \nonumber \\
&& \cdot \sin\left[\omega_i(t-\tau_k)+\angle M_i\right]
+ \nonumber \\
&& +\sum_{k=1}^K|C_k|\sin(\psi_k)\cdot J_0\left[2|M_i|\sin(\omega_i \tau_k)\right]
\label{eq:gamma_thetai_1}
\end{eqnarray}
The first addend of $\Gamma_{\theta}(t)$, reported in the first two lines of \eqref{eq:gamma_thetai_1}, contributes to the system at angular frequency $\omega_i$ in \eqref{eq:system} and can be written as:
\begin{eqnarray}
&& \sum_{k=1}^K|C_k| \cos(\psi_k)\cdot \frac{2 J_1(X_i)}{X_i}
\cdot \left[1-e^{-j2\omega_i\tau_k} \right] \cdot \nonumber \\
&&  \cdot M_ie^{j\omega_i t}= \gamma_{\theta,i} \cdot  M_ie^{j\omega_i t}
\label{eq:gamma_theta_2}
\end{eqnarray}
and again, from \eqref{eq:gamma_theta_2} the expression of $\gamma_{\theta, i}$ can be recognized.
\bibliographystyle{IEEEtran}
\bibliography{Biblio_Rayleigh_Feedback_REVISED}
\end{document}